\documentclass[preprint,12pt,nofootinbib]{revtex4-1}
\usepackage[paper=letterpaper,margin=1.5in]{geometry}
\usepackage[active]{srcltx} 
\usepackage{amssymb,amsmath,amsfonts}
\usepackage{graphicx}
\usepackage{subfigure}

\newcommand{\be}{\begin{equation}}
\newcommand{\ee}{\end{equation}}

\begin{document}

\begin{flushright}
{\normalsize
SLAC-PUB-15932\\
LCLS-II-TN-13-04\\
March 2014}
\end{flushright}

\vspace{.8cm}

\title{Some wakefield effects in the superconducting\\ RF cavities of LCLS-II}

\author{K. Bane\footnote[1]{from SLAC National Accelerator Laboratory}, A. Romanenko\footnote[2]{from Fermi National Laboratory}, and V. Yakovlev\footnotemark[2]}

\maketitle

\section*{Introduction}
Due to the very short longitudinal bunch length, the LCLS-II beam current spectrum extends into the THz range. This means that, as the bunches traverse the superconducting RF (SRF) cavities, some sizable fraction of their wakefield energy is radiated into modes that are above the cut-off frequency and are not trapped.   In the LCLS-II the main source for such radiation is the irises of the RF cavities. However, the transitions at the ends of linacs L1, L2, L3, also are sources of THz radiation loss.
The terahertz radiation is a source of dissipation in SRF cavities that is in addition to the losses of the fundamental mode, which can lead to extra wall heating  and to Cooper pair breaking \cite{1:eq,2:eq}. 

In this note, for LCLS-II, we estimate the power of the radiated wakefields generated in the SRF cavities (including the 3rd harmonic cavities) and in the end transitions. Much of this power will pass through and reflect in the strings of cryomodules that comprise L1, L2, or L3. Presumably, some of it will be absorbed by the higher order mode (HOM) couplers, or by the absorbers at warmer temperatures situated between the cryomodules. In this note, however, we investigate where such power gets generated, but not where it ends up. As such the results can serve as a pessimistic calculation of the extra power that needs to be removed by the cryosystem.

Finally, in this note we also estimate---under the assumption that all the wake power ends up in the SRF walls---the wall heating and the extent of Cooper pair breaking in L3, where the bunch is most intense. Note that, in this note, all calculations are of single bunch effects; thus resonant interactions are not included.

In our calculations we consider the LCLS-II parameters given in Table I. We assume 1.2~MW of beam power, with charge $q = 300$~pC and repetition rate $f_{rep}=1$~MHz. The bunch shape is approximately Gaussian in L1 and L2, and uniform in L3, with rms bunch length $\sigma_z=1000$, 270, 25~$\mu$m in, respectively, L1, L2, L3. 

\begin{table}[hbt]
   \centering
   \caption{LCLS-II parameters used in our calculations (for L3). The longitudinal bunch distribution is approximately uniform. Note that in L1, L2, the bunch distribution is approximately Gaussian, with rms bunch length $\sigma_z=1$, 0.25~mm, respectively.}
   \begin{tabular}{||l|c|c||}\hline 
        {Parameter name} & {Value}  &  Unit\\ \hline\hline 
       Charge per bunch, $q$       &300  &pC  \\
       Rms bunch length, $\sigma_z$       &25  &$\mu$m \\
        Cavity aperture, $a$       &3.5  &cm \\
      Repetition rate, $f_{rep}$       &1  &MHz \\
       Wall dissipation in fundamental        &  & \\       
 \quad \quad mode, per cavity, $P_{diss}$       &13  &W \\
      Time duration of bunch, $\tau=2\sqrt{3}\sigma_z/c$       &290 &fs \\     
Electron-phonon relax. time, $\tau_{e-ph}$       &400 &fs \\
         \hline \hline 
   \end{tabular}
   \label{table1_tab}
\end{table}

\section*{Wakefield losses in cryomodules}

A TESLA cryomodule comprises eight 9-cell cavities, each with active length 1.036~m, and has iris radius $a=3.5$~cm. Between the cavities are bellows that are roughly 6~cm long and have 11 oscillations.
When the beam enters the first cryomodule in a string, it will first encounter transient wakefields that will gradually change to the steady-state wakes. The change occurs over a distance on the order of the catch-up distance,  $z_{cu}= a^2/2\sigma_z$ ($a$ is the iris radius). For LCLS-II, the catch-up distance $z_{cu}= 0.6$, 2.3, 25~m in the three regions. For all three regions $z_{cu}$ is small compared to the length of the sequence of cryomodules, meaning that the steady-state results are a good approximation of the average cryomodule wakes. However, the transient wakes excited in the first cavities (of a sequence) are stronger than the steady-state ones, and need to be considered separately.

\subsection*{Steady-State Wakes}

The steady-state wake effect of the cryomodules depends weakly on bunch length. For sufficiently short bunches the loss factor for a length $L$ of any cylindrically symmetric, periodic structure of iris radius $a$ is given by the asymptotic value
\begin{equation}
\varkappa_{asym}=\frac{Z_0cL}{2\pi a^2}\ ,
\end{equation}
with $Z_0=377$~$\Omega$ and $c$ the speed of light.
Using the asymptotic loss factor (taking $L=L_c=8.3$~m and $a=3.5$~cm) we find the asymptotic value of steady-state power radiated per cryomodule $P_{asym}=11$~W.
A more accurate calculation starts with the point charge wake of a cryomodule, which includes the effects of bellows and pipes between the cavities. The results can be approximated by the simple formula: $W(s)=344.\, e^{-\sqrt{s/s_0}}$ [V/(pC*cryomodule)], where $s_0=1.74$~mm \cite{TESLA}. The loss factor for a Gaussian bunch is given by
\begin{equation}
\varkappa=\frac{1}{2\sqrt{\pi}\sigma_z}\int_0^\infty W(s)e^{-(s/\sigma_z)^2/4}\,ds\ .\label{kappa_eq}
\end{equation}
Using this formula, we find that $\varkappa=86$, 119, 154 V/pC is lost by the beam per cryomodule of, respectively, L1, L2 and L3.
The power lost is then
\begin{equation} 
P_{wake}=q^2\varkappa f_{rep}\ ,
\end{equation}
with $q$ the bunch charge and $f_{rep}$ the repetition rate. With $q=300$~pC and $f_{rep}=1$~MHz, we find that
a total 7.7, 10.7, 13.8 W of power is lost by the beam per cryomodule of, respectively, L1, L2 and L3. 

Note that for a uniform bunch distribution, as is found in L3, if we take $\sigma_z$ to represent the rms length, the loss factor will differ from that given by Eq.~\ref{kappa_eq}, but only by a small amount. This feature is found also in following results that depend on the bunch shape; so rather than present below the equations for uniform distributions we will stick with the more familiar Gaussian ones.
 
\subsection*{Transient Wakes}

For a short bunch, the wake in the first cavity will depend on the history of the bunch's trajectory in the vacuum chamber. However, for simplicity, we here---as is usually done---take as initial condition that the bunch has been in the incoming beam pipe of the first cryomodule for a very long time (longer than a catch-up distance).
 When a short bunch then enters the first cell of the first cavity, the wake induced will be well approximated by the diffraction model, and in subsequent cells and cavities the wake will gradually reach its steady-state form. Let us in this section begin by considering the bunch at it is shortest, in L3, where $\sigma_z=25$~$\mu$m. 

According to the diffraction model, the loss factor for a Gaussian bunch passing through the first cell of a cavity is given by \cite{AChao} 
\begin{equation}
\varkappa=0.723\frac{Z_0c}{\sqrt{2}\pi^2a}\sqrt{\frac{g}{\sigma_z}}\ ,\label{diff_kloss_eq}
\end{equation}
with $g$ the cell gap. For the TESLA cryomodules, the cell period $p=10.5$~cm, and the gap can be taken to be $g=8.9$~cm. Using Eq.~\ref{diff_kloss_eq}, we find that $\varkappa=10$~V/pC is the contribution for the first cell; for the first cavity, the diffraction model would give this value multiplied by the number of cells in a cavity: $\bar\varkappa=90$ V/pC (we will use the bar over $\varkappa$ to indicate loss per cavity). To estimate the loss in cryomodule $n$, $\varkappa_n$, we consider the model
\begin{equation}
\varkappa_{n}\equiv\sum_{m=1}^8\bar\varkappa_{nm}=\frac{1}{9}\sum_{m=1}^8\sum_{p=1}^9\left[ \bar\varkappa_{tr}e^{-\alpha_{nmp}} + \bar\varkappa_{ss}(1-e^{-\alpha_{nmp}})\right]\ ,\label{tr_wake_model_eq}
\end{equation}
with $m$ the cavity number and $p$ the cell number; with $\bar\varkappa_{tr}$, $\bar\varkappa_{ss}$, respectively the transient and steady-state per-cavity loss factor; with $\alpha_{nmp}=[72(n-1)+9(m-1)+p-1]/9d_c$, where $d_c$ is the declination, per cavity, of the transient component.

\begin{figure}[htb]
\centering
\includegraphics*[height=73mm]{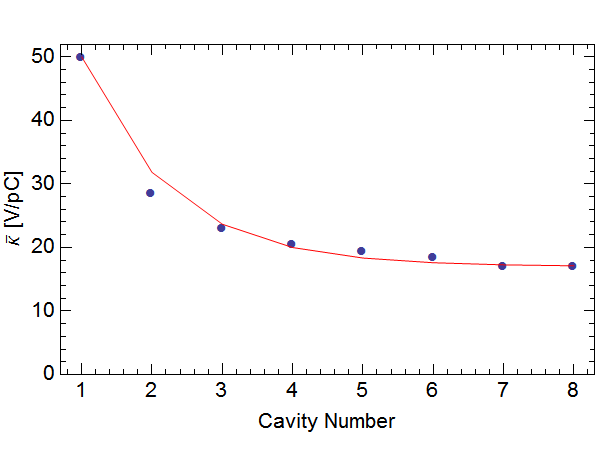}
\caption{For a $\sigma_z=50$~$\mu$m long Gaussian bunch, the loss per cavity $\bar\kappa$ in the first cryomodule. (plotting symbols; taken from Fig.~11 in Ref.~\cite{NovokhatskiMosnier}). The red curve gives the eight terms of $\varkappa_1$ (connected by straight lines) with $d_c=1.25$ (see Eq.~\ref{tr_wake_model_eq}). }
\label{SashaN_kappa_fi}
\end{figure}

Novokhatski and Mosnier have calculated the per cavity loss factor for a $\sigma_z=50$~$\mu$m bunch passing through one TESLA cryomodule (see Ref.~\cite{NovokhatskiMosnier}, Fig.~11). Fig.~\ref{SashaN_kappa_fi}  gives their results (the plotting symbols) and also the 8 terms of $\varkappa_1$ (see Eq.~\ref{tr_wake_model_eq}), with $\bar\varkappa_{tr}=63.6$~V/pC,  $\bar\varkappa_{ss}=17.0$~V/pC,  and $d_c=1.25$ (in red, connected by straight lines). 
The transient loss factor $\bar\varkappa_{tr}$ is given by the diffraction formula, Eq.~\ref{diff_kloss_eq}, with $\sigma_z=50$~$\mu$m, and then multiplied by 9, the number of cells in a cavity; $\bar\varkappa_{ss}$ is taken from Ref.~\cite{NovokhatskiMosnier}, and it agrees well with the steady-state wake formula we used in the previous section. Note that the declination $d_c$ is equivalent to a distance of 1.25~m, which is much less than the so-called catch-up distance, $z_{cu}=a^2/2\sigma_z=12$~m (with $a=3.5$~cm), the distance after which the wake experienced by the beam is within a few percent of the steady-state wake (see {\it e.g.} Ref.~\cite{BaneWeilandTimm}). 
From Fig.~\ref{SashaN_kappa_fi} we see that the agreement between the Novokhatski and Mosnier numerical calculations and our model equation, Eq.~\ref{tr_wake_model_eq}, is good.

We repeat the calculation for the case of L3 in LCLS-II, where $\sigma_z=25$~$\mu$m,  taking $\bar\varkappa_{tr}=90$~V/pC (from the diffraction model), $\bar\varkappa_{ss}=19$~V/pC (from the steady-state section above), and $d_c=2.5$ (since the bunch has half the length of the previous case). We obtain the result that for the first 4 cryomodules the loss factor is: 327, 161, 154, 154~V/pC. For the 300~pC bunch at 1~MHz this corresponds to: 29.5, 14.5, 13.8, 13.8~W power lost by the beam in the first 4 cryomodules.

For completeness, we repeated the calculations for the beam passing through the initial cryomodules of L1 and L2. We find that in L1 the loss in the first cryomodule is 7.8~W, and the result for all the others is 7.7~W; in L2 the loss in the first cryomodule is 11.1~W, and the result for all the others is 10.7~W.

\section*{Wakefield losses due to transitions at ends of linacs}

The optical model of wakefields can be used to estimate the power radiated due to the initial and final transition.  For very short bunches, the optical model depends on bunch length as $\sigma_z^{-1}$.
The average voltage loss (per unit charge) of a short Gaussian bunch passing through a step-out transition, going from a beam pipe of radius $a$ to radius $b$ ($b>a$) is given by the loss factor\cite{HeifetsKheifets,optical}
\begin{equation}
\varkappa= \frac{Z_0c}{2\pi^{3/2}\sigma_z}\ln \frac{b}{a}\ .
\end{equation}
 
For a step-out transition half of the beam loss ($\varkappa/2$) goes into generating primary beam field and half ($\varkappa/2$) is radiated away. However, for a step-in transition (the beam moves from a larger to a smaller pipe), the beam gain in eliminating primary beam field ($\varkappa/2$) equals the amount radiated away ($\varkappa/2$), and the voltage change of the beam itself is near zero \cite{HeifetsKheifets}. Thus, the sum of the average radiated power loss due to the transitions at both ends is given by $\varkappa q^2 f_{rep}$. For the transition from/to a pipe of radius $a = 1$~cm to the cryomodules (with iris radii of $b = 3.5$~cm), it is estimated that 1, 4, 46~W, are radiated into the ends of the cryomodules of, respectively, L1, L2, and L3. The radiation diffracts into the ends of the cryomodules, transversally spreading as the square root of distance; it is estimated that the fields are radiated, respectively, into the first and last 1, 5, 50~m of L1, L2, and L3.

We have treated the wakefield losses due to the transient effects of the cavities and of the transitions at the ends of the linacs separately. In reality the two effects will interfere with each other---reducing the total effect---most noticeable where the bunch is short, in L3. To estimate the real size of the effect in L3, we perform numerical simulations for different bunch lengths, using I.~Zagorodnov's ECHO code, a time domain wakefield solver \cite{Echo}. Ideally we should simulate with an opening and closing transition, with the 20 cryomodules of L3 sandwiched in between. Due to practical considerations, however, the model geometry that we actually use is a step transition from a pipe with 1~cm radius to one with 3.5~cm radius and of length 50~cm; this is followed by the geometry of two cryomodules; at the end is the inverse transition from 3.5~cm to 1~cm. 

The results are shown in Table~II.  Given are the ECHO results of loss factor for two cryomodules $\varkappa_{cr}$, and loss factor for two cryomodules plus two transitions $\varkappa_{tot}$; the (analytic) loss factor of two transitions with large separation $\varkappa_{tr}$, and the ratio $(\varkappa_{tot}-\varkappa_{cr})/\varkappa_{tr}$. The ECHO runs for the shorter bunches were time consuming:  one run for $\sigma_z=50$~$\mu$m took 20~hours on a Windows machine (with a 2.8~GHz Intel Xeon CPU). Extrapolating to  $\sigma_z=25$~$\mu$m, we estimate  $(\varkappa_{tot}-\varkappa_{cr})/\varkappa_{tr}=0.14$. This implies that for L3, because of interference, the extra radiated power due to the end transitions is reduced from 46~W to less than 10~W; and this total amount is distributed over the first and last 50~m of L3.

\begin{table}[hbt]
   \centering
   \caption{ECHO results: bunch length $\sigma_z$, loss factor for two cryomodules $\varkappa_{cr}$, and for two cryomodules plus two transitions $\varkappa_{tot}$. Also shown are the (analytic) loss factor of two transitions with large separation $\varkappa_{tr}$ and the ratio $(\varkappa_{tot}-\varkappa_{cr})/\varkappa_{tr}$. Note: all loss factors are given in units of [V/pC].}
   \begin{tabular}{||c|c|c|c|c||}\hline 
      $\sigma_z$ [$\mu$m]  & $\varkappa_{cr}$ & $\varkappa_{tot}$  & $\varkappa_{tr}$ & $(\varkappa_{tot}-\varkappa_{cr})/\varkappa_{tr}$\\ \hline\hline 
1000 & 175. &189.  & 13. &1.10 \\
400 & 222. &254. & 32. & 1.01\\
200 & 257. &306.  & 64. &0.77 \\
100 & 295. & 401. & 127. & 0.83\\
50 &343.  & 448. & 254. & 0.41\\
         \hline \hline 
   \end{tabular}
   \label{table1_tab}
\end{table}

\section*{Wakefield Losses in the 3.9~GHz Cryomodules}

Two 3.9~GHz SRF cryomodules will be installed upstream of BC1 for longitudinal phase space control. The total length of each cryomodule is 12~m. Each cryomodule comprises eight 9-cell cavities, each of which has active length $L_{cav}=34.6$~cm; the cavity-to-cavity spacing is 1.38~m. The iris radius $a=1.5$~cm. 

Many details of the cryomodule layout have not yet been set. Rather than attempt a simulation of the wake at this point, we will just make an estimate of the power generated by the beam passing through the 3.9~GHz cryomodule. I.~Zagorodnov et al have performed detailed calculations for the 3.9~GHz cryomodules to be used in X-FEL~\cite{Zagorodnov_3p9}. The X-FEL 3.9~GHz cryomodule has the same cavity shape as will be used in LCLS-II. However, each 3.9~GHz cryomodule of X-FEL has only 4 cavities (plus bellows and end transitions).The authors find that, for a $\sigma_z=1$~mm bunch, the per cryomodule loss factor is $\varkappa=71$~V/pC.  For an estimate for LCLS-II, with its 8 cavities per cryomodule, we simply multiply their result by two: {\it i.e.} we let $\varkappa=142$~V/pC. Then, with $q=300$~pC and $f_{rep}=1$~MHz, the power radiated by the beam in each cryomodule is estimated to be $13$~W. In the future, when the LCLS-II cryomodule layout is set, numerical simulations should be performed to confirm this estimate.

\section*{Pulsed temperature rise caused by the bunch fields}

The effects we consider in this and following sections are most pronounced when the beam has high peak current, and since the bunches have the highest peak current in L3, from here forward we will limit ourselves to considering only the L3 cryomodules; all the analyzed effects will only be weaker in L1 and L2. 

In L3 the bunch shape is approximately uniform, and the instantaneous current during a pulse is $I=qc/(2\sqrt{3}\sigma_z)= 1$~kA, which produces magnetic field of the amplitude $H=I/(2\pi a)\approx 4.6$~kA/m on the surface of the aperture. Since $\tau\ll \tau_{e-ph}$ there is no effective Meissner screening and this will lead to the instantaneous dissipation power of about $P_d\approx \rho  H^2/(2\ell)$ where $\rho\sim1$~n$\Omega\cdot$m is the normal state electrical resistivity at 2K, and $\ell\sim$1~$\mu$m is an electron mean free path in high RRR niobium. We obtain $P_d\approx1$~W/cm$^2$ during time of order $\tau$ leading to the energy deposition
 per unit volume of $ \Delta W/\Delta V=P_d\tau/\ell=3.1$~nJ/cm$^3$.
 Taking the specific heat of superconducting Nb at 2K \cite{4:eq} to be $c_{heat}=0.12$~mJ/(cm$^3\cdot$K), we obtain for the pulse heating $\Delta T_{pulse}=(\Delta W/\Delta V)/c_{heat} \le0.025$~mK. 

\section*{Average temperature rise caused by the bunch fields}

The time-averaged dissipated power is $P_d \tau f_{rep}\approx3.1$~mW/m$^2$. Taking a niobium wall thickness of 3~mm, the thermal conductivity and Kapitza resistance from \cite{5:eq}, and solving for the steady state heat diffusion, we find that there will be a negligible temperature increase on the cavity wall (near the aperture), $\Delta T_{avg}\approx0.004$~mK. Thus, neither thermal quench nor extra dissipation---due to non-equilibrium Meissner screening around the apertures---are issues. If we take the affected area to be of width $d\sim1$~cm around each aperture, this will lead, for a 9-cell cavity with 10 apertures, to a deposited energy of about $10 \times P_d 2\pi ad\,\tau\approx0.07$~nJ per bunch, or an additional time-averaged power of $P_{davg}=0.07$~mW$\ll P_{diss}$. It is important to note that the lack of Meissner screening of the magnetic field for the ultrafast bunch is purely a non-equilibrium, relaxation effect which does not directly affect the superconducting surface resistance and thus the dissipation in the fundamental mode. An additional dissipation in the beam pipe of length $\sim13$~cm will be $P_d 2\pi a\tau f_{rep} \times13$~cm$=0.08$~mW, which is small compared to the thermal flow from the beam pipe and coupler, $\sim0.12$--0.16 W (for the end cavities).

\section*{Estimate of the dissipation caused by wakefields}

We now turn to wakefields left behind in the cavity. The loss factor for a sequence of TESLA type cryomodules (eight 9-cell cavities in each) was numerically evaluated in \cite{TESLA} for somewhat longer bunches ($\sigma_z\ge 50$~$\mu$m). For the bunch length of $\sigma_z=25$~$\mu$m we have transient (steady-state) loss factors $\varkappa=720$ (150)~V/pC (see above), wake energy loss $U_{wake}=q^2\varkappa=30$ (14) $\mu$J, and time averaged deposited power is $P_{wake}=30$ (13.8)~W; all these results are per cryomodule. These results are small compared to $8P_{diss}=108$~W, the wall dissipation in the fundamental mode (per cryomodule). As was shown in \cite{1:eq} all the photons in the wakefield pulse are randomly reflected many times before their eventual absorption in the SRF cavity walls;  the characteristic timescale is hundreds of nanoseconds, which is comparable to the inter-bunch spacing of 1~$\mu$s for LCLS-II. Therefore, wakefield energy deposition is essentially homogeneous over the whole cavity surface and the instantaneous power absorbed is close to the time-averaged power absorbed.

\section*{Cooper pair breaking by THz radiation}

The additional power $P_{wake}$ will increase wall losses in the fundamental mode due to two effects: an increase in RF surface temperature $\Delta T$, and an increase in the fraction of unpaired electrons $\Delta n_N$. 
An estimate of $\Delta T$ using the same parameters as above gives  $\Delta T\cong1$~mK, and a corresponding additional dissipated power $P_1\cong0.1$~W$\ll P_{diss}$. 

To estimate the extent of the breaking of Cooper pairs in the niobium by the fields of the beam, we calculate the wakefield power for frequencies above the pair breaking threshold frequency, $f_{cpb}=750$~GHz. When the beam traverses the beginning of L3, the high frequency impedance is one that can be approximated by the diffraction model; eventually, the high frequency impedance of a periodic structure applies. Of the two models, the diffraction model power drops slower at high frequencies, so it is in the first cavities of L3 that the breaking of Cooper pairs will be largest in number.

The relative power radiated above the Cooper pair breaking threshold can be approximated by
\begin{equation}
r_{cpb}=\frac{1}{2\pi\varkappa}\int_{f_{cpb}}^\infty R(\omega)e^{-(\omega\sigma_z/c)^2}d\omega\ ,\label{r_eq}
\end{equation}
with $R(\omega)$ the real part of the impedance and $\omega$ the frequency.
For the transient wake we use the diffraction model \cite{AChao}
\begin{equation}
R(\omega)=\frac{Z_0M}{2\pi^{3/2}a}\sqrt{\frac{cg}{\omega}}\ ,
\end{equation}
with $M$ the number of cells, $a$ the radius of the beam pipe, and $g$ the gap of the cavity cells. For the transient wake we obtain $r_{cpb}=0.33$.

For the steady-state wake we use the periodic diffraction model of Gluckstern \cite{Gluckstern,YokoyaBane}
\begin{equation}
Z(\omega)=\frac{iZ_0cL}{\pi \omega a^2}\left[1+\frac{\alpha(g/p)p}{a}\sqrt{\frac{2\pi i c}{\omega g}}\right]^{-1}\ ,
\end{equation}
with $Z(\omega)$ the impedance, $L$ the length of the structure, $a$ the radius of the irises, $p$ the period, and $g$ the gap of the cavity cells; with $\alpha(\zeta)=1-0.465\sqrt{\zeta}-0.070\zeta$. At high frequencies $R(\omega)$ for this model drops as $\omega^{-3/2}$, which is faster than the $\omega^{-1/2}$ dependence for the diffraction model.
Numerically performing the integral Eq.~\ref{r_eq}, we find that for the steady-state wake $r_{cpb}=0.01$.

The total number of Cooper pairs in the magnetic field penetration depth $\delta\sim100$~nm (where photon absorption happens) of one 9-cell cavity with surface area $S_A=0.8$~m$^2$ at 2K is given by \cite{Cooper}
\begin{equation}
N_{Cooper}\cong\frac{\Delta E}{E_f}n_eS_A\delta\ ,
\end{equation}
with the band gap of niobium $\Delta E=1.55\times10^{-3}$~V, the Fermi energy $E_f=5.35$~V;  where the density of normal conducting electrons $n_e=\rho Z/(A m_p)$, with niobium density $\rho=8.57\times10^5$~kg/m$^3$, atomic number $Z=41$, atomic weight $A=93$, and proton mass $m_p=1.672\times10^{-27}$~kg. We find that $\Delta E/E_f=2.9\times10^{-4}$, $n_e=2.3\times10^{30}$, and finally $N_{Cooper}=5\times10^{19}$.

Converting the total wakefield energy deposited per bunch into number of $f\ge750$~GHz photons (in a cavity; remember there are 8 cavities in a cryomodule) we obtain:
\begin{equation}
N_{ph}=\frac{1}{8}\frac{r_{cpb}U_{wake}}{hf_{cpb}}\ ,
\end{equation}
with $h=6.63\times10^{-34}$~m$^2$kg/s, Planck's constant. We find that, for the transient (steady-state) case, $N_{ph}=5.4$ (0.04) $\times 10^{15}$, which in both cases is negligible compared to $N_{Cooper}$. 
Thus pair-breaking induced by both the increase in normal fluid density and in the surface resistance are negligible. Since the characteristic electron-phonon relaxation time is on order 0.4~ps, by the time the next bunch arrives in 1~$\mu$s, the number of Cooper pairs is back to thermal equilibrium, and no cumulative effects are present.

\section*{Summary}

In this note we calculated the power radiated by the beam that can end up in the cryomodules. We considered the worst case scenario of charge $q=300$~pC and repetition rate $f_{rep}=1$~MHz. From the RF cavities themselves, the steady-state loss is 8, 11, 14~W per cryomodule in the three linacs; the loss in the first cryomodule of L3, however, is a transient that is estimated to be 29~W. For the radiation generated in the 1~cm to 3.5~cm (radius) transitions at the ends of the three linacs, we estimate that 1~W and 4~W (total; {\it i.e.} half at the beginning and half at the end of these linacs) is radiated by the beam in L1 and L2, respectively.  For the case of L3, where the bunch is the shortest, there is interference between this transition wake and the initial wake of the cryomodule that follows. Through a model simulation we estimate that the extra contribution due to the transitions at the ends of L3 is $<10$~W (in total). Finally, the power lost by the beam in each of the two 3.9~GHz cryomodules is estimated to be 13~W.

Since the power lost by the beam $P=q^2\varkappa f_{rep} $, the power results for the nominal $q=100$~pC case will be much reduced compared to these numbers. The loss factor $\varkappa$, in general, depends on bunch length and bunch shape. After BC2 the bunch distribution is (approximately) uniform, with a peak current of $\hat I=1$~kA, implying an rms bunch length $\sigma_z=8.3$~$\mu$m. In this region of LCLS-II with $q=100$~pC, the transition loss  (since $\varkappa\sim\sigma_z^{-1}$ and there will be more interference) is $<1.9$~W; for the cavity loss, the steady-state power (since $\varkappa$ is weakly dependent on $\sigma_z$) is 1.5~W, and the transient estimate for the first cryomodule (since $\varkappa\sim\sigma_z^{-1/2}$) becomes 5.6~W.  

We also estimated the heating and Cooper pair breaking due to the wake, and conclude that wakefield effects on the superconducting SRF cavities in LCLS-II are small---even under the pessimistic assumption that all the radiated power is absorbed in the cavities.

\section*{Acknowledgement}
 We thank G. Stupakov for helpful discussions. Work
supported by Department of Energy contract DE--AC02--76SF00515.

\end{document}